\documentstyle[11pt,aaspp4]{article}
\input psfig
\psfull
\singlespace
\parskip=\the\medskipamount

\def\be{\begin{equation}}
\def\ee{\end{equation}}
\def\bea{\begin{eqnarray}}
\def\eea{\end{eqnarray}}
\def\te{\Delta t_{\rm Earth}}

\def\tf{\Delta t_{\rm frustrated}}
\def\tb{\Delta t_{\rm biogenesis}}
\def\tl{\Delta t_{\rm life}}

\def\tn{\Delta t(N)}

\begin{document}
\title{
Does the Rapid Appearance of Life on Earth\\
 Suggest that Life is Common in the Universe?}
\medskip 

\author{Charles H. Lineweaver, Tamara M. Davis\\
School of Physics, University of New South Wales\\
%Sydney, NSW 2052,Australia \\
%tel: 61-2-9385-5168\\
%fax  61-2-9385-6060\\
charley@bat.phys.unsw.edu.au}
%Running Header: ``Does the Rapid Appearance...''\\
%Keywords: Exobiology, terrestrial planets, Earth, prebiotic environments, SETI}
%Address editorial correspondence to:\\
%Charles H. Lineweaver\\
%School of Physics, University of New South Wales\\
%Sydney, NSW 2052,Australia \\
%tel: 61-2-9385-5168\\
%fax  61-2-9385-6060\\
%charley@bat.phys.unsw.edu.au\\

%%%%%%%%%%%%%%%%%%%%%%%%%%%%%%%%%%%%%%%%%%%%%%%%%%%%%%%%%%%%%%%%%%%%%%%%%%%%%%
\begin{abstract}
It is sometimes assumed 
%(e.g. Lineweaver 2001) 
that the rapidity of biogenesis on Earth suggests that life
is common in the Universe.
Here we critically examine the assumptions inherent in this 
if-life-evolved-rapidly-life-must-be-common argument.
We use the observational constraints on the rapidity of 
biogenesis on Earth to infer the probability of biogenesis 
on terrestrial planets with the same unknown probability
of biogenesis as the Earth.
We find that on such planets, older than $\sim 1$ Gyr, 
the probability of biogenesis is $> 13\%$ at the $95\%$ 
confidence level.
This quantifies an important term in the Drake Equation
but does not necessarily mean that life is common in 
the Universe.
\end{abstract}

%\keywords{Terrestrial Planets, Extrasolar Planets, Cosmochemistry, Planetary 
%Formation, Planets, General}

%%%%%%%%%%%%%%%%%%%%%%%%%%%%%%%%%%%%%%%%%%%%%%%%%%%%%%%%%%%%%%%%%%%%%%%%%%%%%%
%\clearpage
\section{The Biogenesis Lottery}
\label{sec:intro}
Much of current astrobiological research is focused on learning
more about the early evolution of the Earth and about the
origin of life. We may be able to
extrapolate and generalize our knowledge of how life formed here 
to how it might have formed elsewhere.
Indirect evidence suggesting that life may be common in the Universe
includes:
\begin{itemize}
\item Sun-like stars are common.
\item Formation of Earth-like planets in habitable zones around these stars 
may be a common feature of star formation (Kasting et al. 1993, 
Wetherill 1996, Lissauer and Lin 2000, Lineweaver 2001).
\item Life's chemical ingredients -- water, amino acids and other 
organic molecules -- are common 
(Cronin 1989, Trimble 1997, Charnley et al. 2002).
\item  Sources of free energy such as starlight and reduction-oxidation pairs
are common (Nealson and Conrad 1999).
\end{itemize}
\noindent It is difficult to translate this circumstantial 
evidence into an estimate of how common life is in the Universe.
Without definitive detections of extraterrestrial life we can 
say very little about how common it is or 
even whether it exists.
Our existence on Earth can tell us little about 
how common life is in the Universe or about the probability 
of biogenesis on a terrestrial planet because, 
even if this probability were infinitesimally small and 
there were only one life-harboring planet in the Universe 
we would, of necessity, find ourselves on that planet.
However, the rapidity with which life appeared on Earth gives
us more information.
If life were rare it would be unlikely that
biogenesis would occur as rapidly as it seems to have occurred
on Earth.

Although we do not understand 
the details of how life originated,
we have some useful observational constraints on how long
it took.
Carbon isotopic evidence suggests that life existed on Earth more than 
$3.85$ billion years ago (Mojzsis et al. 1996).
High temperatures and large frequent sterilizing 
impacts may have frustrated an earlier appearance of life 
(Maher and Stevenson 1988, Sleep et al. 1989). 
If life originated on Earth, then increasingly tight observational
constraints indicate that biogenesis was rapid 
(Oberbeck and Fogleman 1989, Sleep et al. 2001). 
The extraterrestrial implications of rapid biogenesis on Earth and 
the extent to which this rapidity suggests that life is
common in the Universe have not been looked at carefully and are
the focus of this paper.

The basic concept is simple; over a given time period, more probable 
events happen more often (and thus more rapidly) than less probable events. 
Thus, the probability of winning a lottery can be inferred from how quickly
a lottery winner has won.
For example, suppose we have no idea about the probability $q$, of winning a 
daily lottery ($0 \leq q \leq 1$). Suppose a gambler buys a lottery 
ticket every day for three days, losing on the first two days and winning
on the third. We can use this information to infer 
something about $q$. Specifically, in this case, we can say that $q$ is 
more likely to be about one third than one 
hundredth, and is unlikely to be close to 1 (see ${\mathcal{L}}(n=3;q)$ in Fig. 1). 
If the gambler can only tell us that he won at least once within 3 days
then we can no longer exclude high values of $q$ with such confidence, 
but the likelihood of $q$  can still tell us that $q > 0.16$ 
at the $95\%$ confidence level (see ${\mathcal{L}}(\leq 3;q)$ in Fig. 1 and Eq. \ref{eq:A2} of Appendix).

Suppose there is a group of gamblers, all of whom have won 
at least once within $N$ (e.g. $12$) days. A gambler is choosen 
from this group at 
random and after carefully examining his tickets, he tells us that 
he won at least once within the first 3 days --
relatively early in the 12 days that he had to have won by, to be in the group.
This is analogous to our situation on Earth. We find ourselves 
in the group of planets on which biogenesis has necessarily 
occurred -- we have of necessity won the biogenetic lottery some time 
in the past. And we also find that biogenesis has occurred rapidly.
We won soon after life became possible on Earth.
Given the above information about the gambler, the likelihood of $q$ is 
plotted in Fig. 1 as ${\mathcal{L}}(n\leq 3, N \leq 12 ;q)$ and allows us to
conclude that $q>0.12$ at the $95\%$ confidence level.
This statistical result applies to a group of gamblers or to a group of terrestrial planets 
on which the probability $q$ of biogenesis is unknown, as long as $q$ is 
approximately the same for each planet in the group and 
approximately the same as it was on Earth.
In the next section we review the observational constraints on 
when and how quickly life appeared on Earth.
In Section \ref{sec:caveats} we use these constraints to identify and 
critically examine selection effects that complicate this result.
In Section \ref{sec:Drake} we discuss the relationship between our 
result, the Drake Equation, and the larger question: 
`How common is life in the Universe?'
Mathematical details %of the probability inferences 
are relegated to the appendix.

%%%%%%%%%%%%%%%%%%%%%%%%%%%%%%%%%%%%%%%%%%%%%%%%%%%%%%%%%%%%%%%%%%%%%%%%%
%%%%%\clearpage
%\vspace{15cm}
\begin{figure*}[!h!t]
\centerline{\psfig{figure=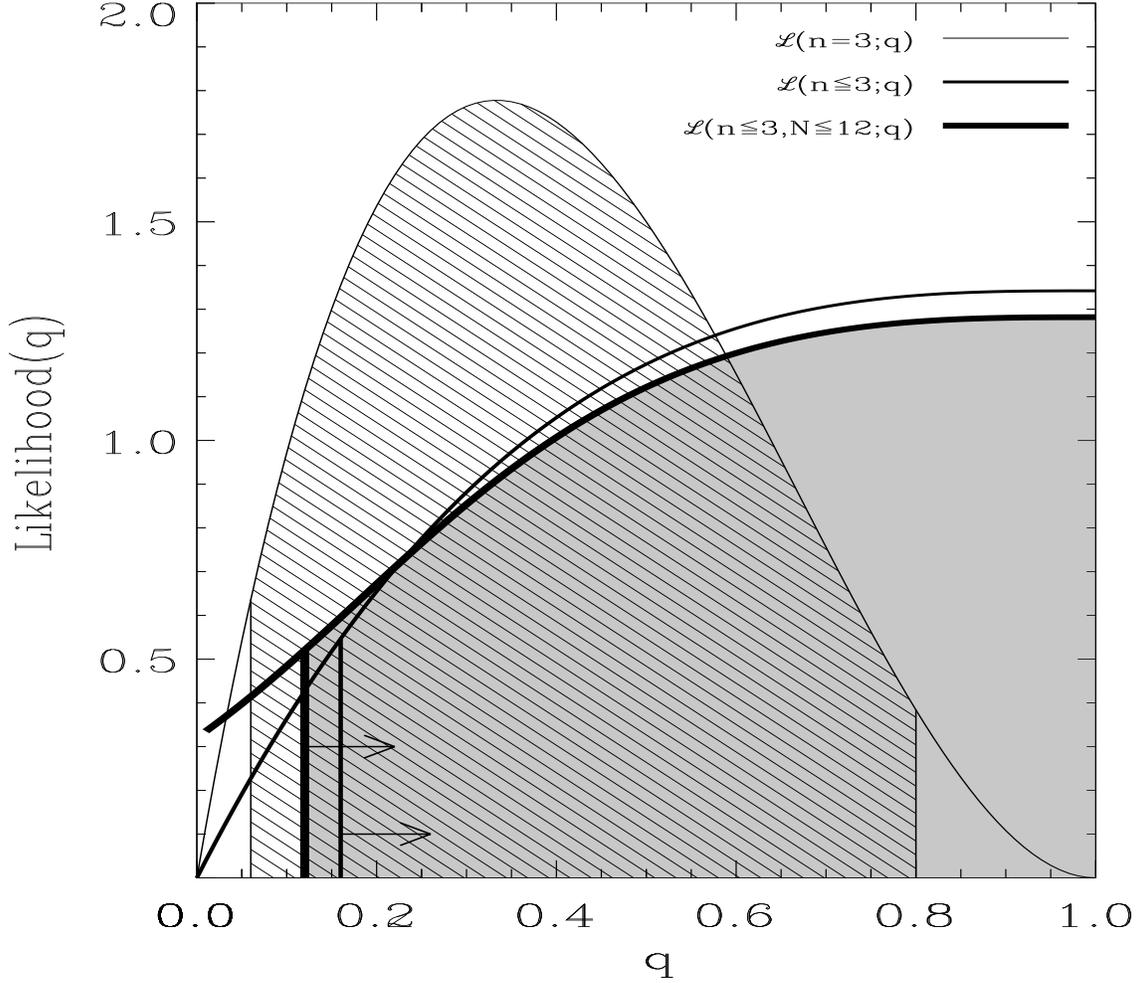,height=13.0cm,width=15.0cm}}
\caption{Let the unknown probability of winning a daily lottery be $q$.
Suppose a gambler buys a ticket each day and wins on the third day. 
From this information we can calculate the likelihood of $q$
(see ${\mathcal{L}}(n = 3;q)$ above).
The most likely value is, as expected $1/3$, but we can also conclude 
that  $0.06 < q < 0.80$ at the 95\% confidence
level (hatched area).
Different scenario: suppose that after three days the gambler tells us that 
he won at least once. The likelihood of $q$ then becomes 
${\mathcal{L}}(n \leq 3; q)$ plotted above.
High values of $q$ can no longer be excluded and we can say only that
$0.16 < q $ at the 95\% confidence level.
Different scenario (and one more analogous to our situation on Earth): 
we have a group of gamblers, all of whom have won 
at least once, on or before the $N$th (e.g. $12$th) day. One of them, 
chosen at random (analogous to the Earth), won at least once, on or before the $3$rd day. 
This is quite early since it could have happened any time during 
the 12 days. Given this 
information, the likelihood of $q$ becomes the thick curve labelled 
`${\mathcal{L}}(n \leq 3, N \leq 12;q)$'. In this case, we can say 
that $q > 0.12$ at the 95\% confidence level (grey area). 
%The dotted curves show the effect of changing $N$. 
See Appendix for computational details. 
}
\label{fig:lottery3}
\end{figure*}
%%%%%%%%%%%%%%%%%%%%%%%%%%%%%%%%%%%%%%%%%%%%%%%%%

%%%%%%%\clearpage
\section{Observational Constraints on the Timing of Terrestrial Biogenesis}
\label{sec:observations}

If life originated on Earth, then during and immediately following the Earth's 
formation there was a period  without life ($\tf$),  
followed by a period during which life evolved ($\tb$), 
followed by a period during which life has been present ($\tl$).
The sum of these intervals adds up to the age of the Earth (see Fig. 2):
\be
\tf + \tb + \tl = \te
\label{eq:simple}
\ee
where $\te = 4.566 \pm 0.002$ Gyr (All\`{e}gre et al. 1995).
As older fossils and biosignatures have been found, $\tl$ has gotten longer.
The significance of large impacts in frustrating or sterilizing proto-life has 
only recently been appreciated and assessed ($\tf$).
Combined, these observations indicate that biogenesis was rapid since
$\tb$ is caught in the middle -- the longer $\tf$ and $\tl$ get, the shorter
$\tb$ must get.
The distinction between how rapid biogenesis was and when it
was, is important because our result, the inferred 
probability of biogenesis, depends on how rapid it was, while only a 
marginal selection effect depends on when it
was (Section \ref{sec:caveats}).

The majority of the Earth's mass accreted from planetesimals within the 
first 100 million years of the Earth's formation %($\sim 4.55$ to $\sim 4.45$ Gyr ago 
(Halliday 2000). 
With an initially molten surface, life could not have appeared.
The transition from accretion to heavy bombardment included the formation of the 
Moon  by the collision with a Mars-sized object $\sim 4.5$ Gyr ago 
(Hartmann and Davis 1975, %Newsom \& Taylor 1988, 
Halliday 2001, Canup and Asphaug 2001).  
We can infer from the dates and sizes of lunar impact crators, whose record
goes back to when the Moon formed a solid crust 
($\sim 4.44$ Gyr ago, Sleep et al. 1989) 
that the surface of the Earth was periodically vaporized. 
Since the mass of the Earth is 80 times the mass of the Moon, impacts on 
the Earth were more numerous, more energetic and
periodically produced 2,000 K rock vapor atmospheres which lasted 
for several thousand years (Hartmann et al. 2000, Sleep et al. 2001).
These conditions were probably an effective and recurring autoclave for 
sterilizing the earliest life forms or more generally frustrating the evolution 
of life. A steadily decreasing heavy bombardment continued until 
$\sim 3.8$ Gyr ago.

Estimates of the time of the most recent sterilizing impact range
between $4.44$ and $3.7$ Gyr ago  (Maher and Stevenson 1988, Sleep et al. 1989,
Oberbeck and Fogleman 1989, Halliday 2001).
These estimates span the time from the solidification of the 
Moon's crust to the end of the late heavy bombardment.
Thus, life was frustrated for at least the 
first 0.1 Gyr and possibly as long as the first 0.9 Gyr of the 
Earth's existence.
We take our preferred value as the middle of this range: 
$\tf \approx 0.5 \pm 0.4$ Gyr.
The range of these estimates reflects the large uncertainties due to 
small number statistics for the largest impactors and the uncertainty of 
the energy required to sterilize the Earth completely.
We do not know where biogenesis happened or the extent to 
which it was protected from the effects of impacts.
Tidal pools have little protection, hydrothermal vents 
have some protection, while autotrophic thermophiles in sub-surface rock 
under several kilometers of crust were probably in effective bomb shelters.

The roots of the universal tree of life point to a thermophilic 
origin (or at least a thermophilic common ancestor) for all life on earth
(Pace 1991, Stetter 1996).
%This, as well as the long branch lengths near the roots of the tree
%(Gogarten-Boeckels et al. 1995), 
This suggests a hot biogenesis in hydrothermal vents or 
possibly sub-surface rock and/or selection for thermophilia by 
periodic temperature pulses from large impacts.
${\it{If}}$ we knew that life evolved on the surface of the 
Earth and was therefore more susceptible to impact sterilizations, 
life would have been frustrated longer and our preferred value would 
be more precise: $\tf \approx 0.7 \pm 0.2$ Gyr.

If we accept the carbon isotopic evidence for life more than 
$3.85$ billion years ago (Mojzsis et al. 1996) then life has been on Earth
{\bf at least} that long, i.e., $\tl$ is at least $3.85$ Gyr.
%Schidlowski et al. 1988, Rosings et al. 1999). 
In addition, because of the
Earth's tectonic history, this time is also the earliest time we could 
reasonably hope to find biological evidence from rocks on Earth -- even 
if life existed earlier. With this selection effect in mind (which we 
know exists at some level), our preferred value for the time life has 
existed on Earth is: 
$\tl \approx 4.0^{+0.4}_{-0.2}$ Gyr.

It is possible that biogenesis occurred several times on Earth.
For example, during the period $\tf$, life could 
have evolved and been sterilized multiple times. 
We do not know if this happened. 
%Thus we have no observational constraints on the duration of such periods of biogenesis. 
Similar potential sterilizations and biogeneses could
have occurred during the period $\tl$ but we do not know.
For the purposes of this analysis we can ignore this
complication. We are only interested in the shortest period 
within which the observations 
%(with crude corrections for selection effects)
can constrain biogenesis to have occurred.
Thus, $\tb$ is our best observational constraint
on any epoch of biogenesis and this happens to be
on the period between the most recent sterilizing impact that is older 
than the oldest evidence we have for life on Earth. 

Since our preferred values yield $\tf + \tl \approx \te$, there is 
little time left for biogenesis to
have occurred. This is the basis for the statement that biogenesis occurred
rapidly. Substituting our three preferred values,
$\te,\; \tf$ and $\tl $ into Eq. \ref{eq:simple}
and solving for $\tb$ yields,
\be
\tb= 0.1^{+0.5}_{-0.1}\; {\rm {\mbox Gyr.}}
%\tb= 0^{+0.418}_{-0.636}
\ee
Thus we take $600$ Myr as a crude estimate of the 
upper limit for the time it took life to appear on Earth.
Assuming biogenesis took place on the surface of the Earth, 
Oberbeck and Fogleman (1989) found this maximum time to be $ \sim 25$ Myr.
Maher and Stevenson (1988) assumed that biogenesis took somewhere 
between $0.1$ Myr and $10$ Myr depending on environment, while
Sleep et al. (2001) find a similar range for evolutionarily significant
periods of clement surface conditions.
Independently, biologists specializing in the chemistry of the 
origin of life have estimated that the time required for biogenesis 
is potentially quite short (Miller 1982) and could be less than 
8 Myr (Lazcano and Miller 1994).
Thus, several lines of evidence indicate that biogenesis was 
geologically rapid. 
If biogenesis occurred in the fissures around a hydrothermal vent
or in subsurface rocks, it was well-protected and we can only 
constrain biogenesis to have taken less than $\sim 600$ Myr. 
If biogenesis occurred on the surface, it probably took less 
than $\sim 25$ Myr. 
In this analysis we consider $600$ and $25$ Myr to represent
the high and low ends of a plausible range for $\tb$ 
(see Figs. 2 and 3 respectively).

These observational constraints on $\tf$, $\tb$ and $\tl$ are important because
they quantify how rapidly biogenesis occurred on Earth and enable us to put 
limits on the probability that biogenesis occurred on other planets. 
For example, consider a group
of terrestrial planets with approximately the same probability of biogenesis
`$q$' as Earth. Suppose $q= 0.30$. At their
formation, none of these planets had life. As time passed, life arose on
more and more of them. The thick line in Fig. 2 shows the increasing percentage of these
planets with life as time passes. After $\tb$, $30\%$ will have life (that is how $q=0.30$ is defined).
After $4.566$ Gyr, $93\%$ will have life ($7\%$ still will not).
Of that $93\%$, $32\%$ will, like the Earth, have had biogenesis within $\tb$. 
The histogram on the far right of Fig. 2 represents these numbers.
Suppose $q$ is only $0.03$. Then only $20\%$ will have life after $4.566$ Gyr and only 
$14\%$ of those will have had biogenesis, like the Earth, within $\tb$. 
Assuming Earth is a random member of the planets with life, the single 
observation that biogenesis occurred within $\tb$ on Earth indicates 
that larger values of $q$ are more likely than small values.
This is the basic idea behind our analysis.
It is illustrated in Figs. 1, 4 and 5 which also show quantitative constraints on $q$ under 
the various assumptions discussed in Section 3.

The histograms in Figs. 2 and 3 also show that if $q$ is large, the fraction of 
gamblers or terrestrial planets who have won after a certain time is large.
Suppose the gambler did not know how many gamblers had won by the 12th day, 
but only that he is a random member of the group that had won. 
From the fact that he won quickly, he can infer that $q$ is large.
This tells the gambler that after a few days a large fraction of lottery
ticket buyers are in the winners group. He is not alone.
For the biogenesis lottery, finding large $q$ 
means that after a few times $\tb$,  a large fraction of the 
terrestrial planets with probability of biogenesis similar to the Earth,
have (or have had) life.

%%%%%%%%%%%%%%%%%%%%%%%%%%%%%%%%%%%%%%%%%%%%%%%%%%%%%%%%%%%%%%%%
\clearpage
%\vspace{15cm}
\begin{figure*}[!h!t]
\centerline{\psfig{figure=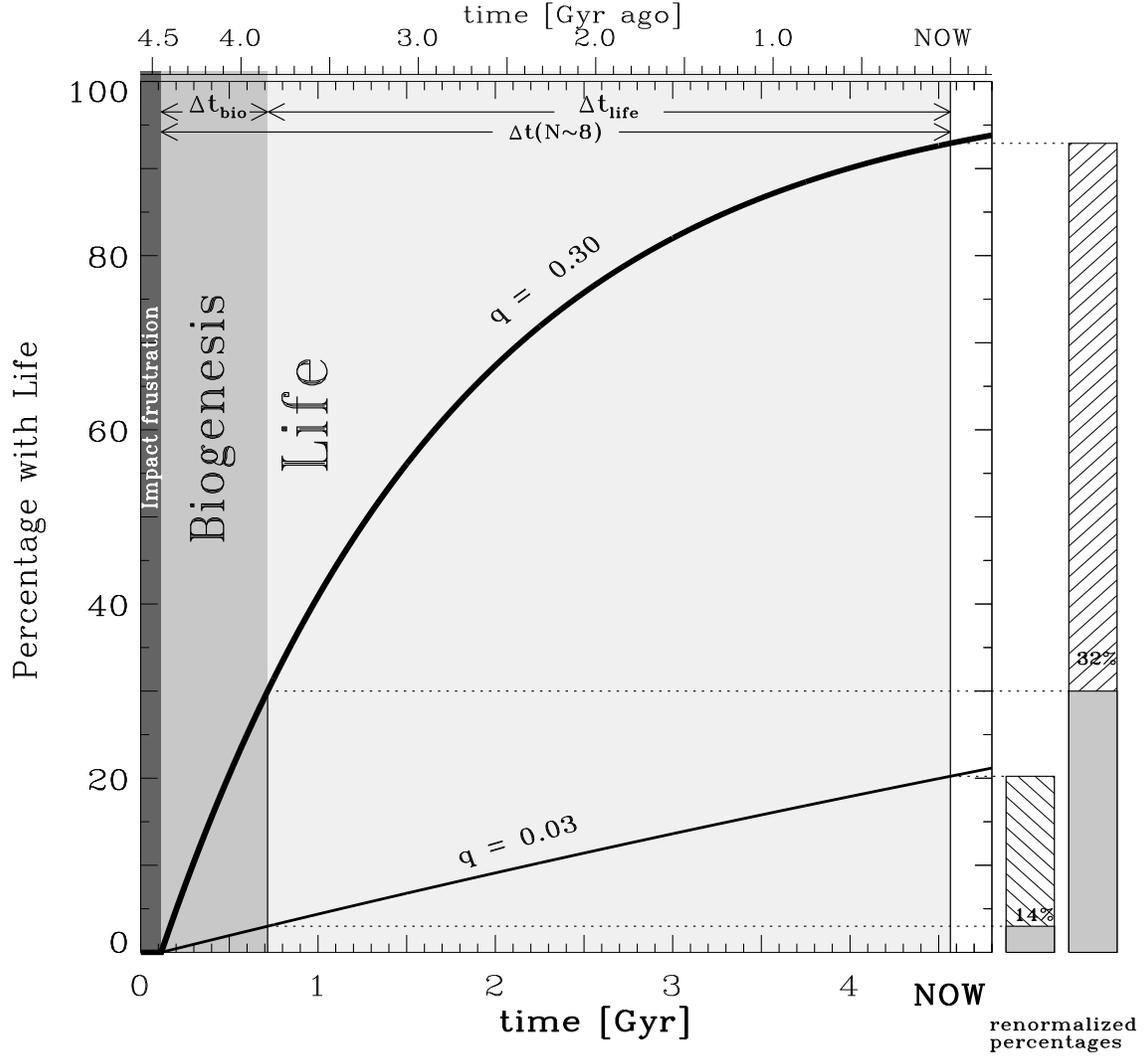,height=14.0cm,width=15.0cm}}
\caption{We divide the history of the Earth into three epochs: 
Impact frustration, Biogenesis and Life. The black curves show 
the percentages of terrestial planets with life as a function of time
assuming two different probabilities of biogenesis ($q=0.3,\; 0.03$),
within $\tb = 600$ Myr.
The percentages in the histograms on the right ($14\%$ and $32\%$) 
are obtained from comparing the subset of planets that have formed life 
within $\tb$ (middle grey) with the total number of 
planets that have life (or have had biogenesis) within $4.566$ Gyr 
of their formation (cross-hatched).
If $q$ is high ($0.30$, thick line) a large fraction ($32\%$) 
of the planets which have evolved life within $4.566$ Gyr of formation,
%(after $t_{terminate} = 4566$ million years)
have life that evolved rapidly -- within $\tb$ --
on their planets. If $q$ is low ($0.03$, thin line) then a 
smaller fraction ($14\%$) will have life that evolved rapidly.
These different percentages illustrate the principle that
a single observation of rapid terrestrial biogenesis
is more likely to be the result of high $q$.
This allows us to compute the relative likelihood of $q$ and 
to constrain $q$ (see Fig. 4).
}
\label{fig:time3}
\end{figure*}
%%%%%%%%%%%%%%%%%%%%%%%%%%%%%%%%%%%%%%%%%%%%%%%%%
%%%%%%%%%%%%%%%%%%%%%%%%%%%%%%%%%%%%%%%%%%%%%%%%%%%%%%%%%%%%%%%%%%%%%%%%%
\clearpage
%\vspace{15cm}
\begin{figure*}[!h!t]
\centerline{\psfig{figure=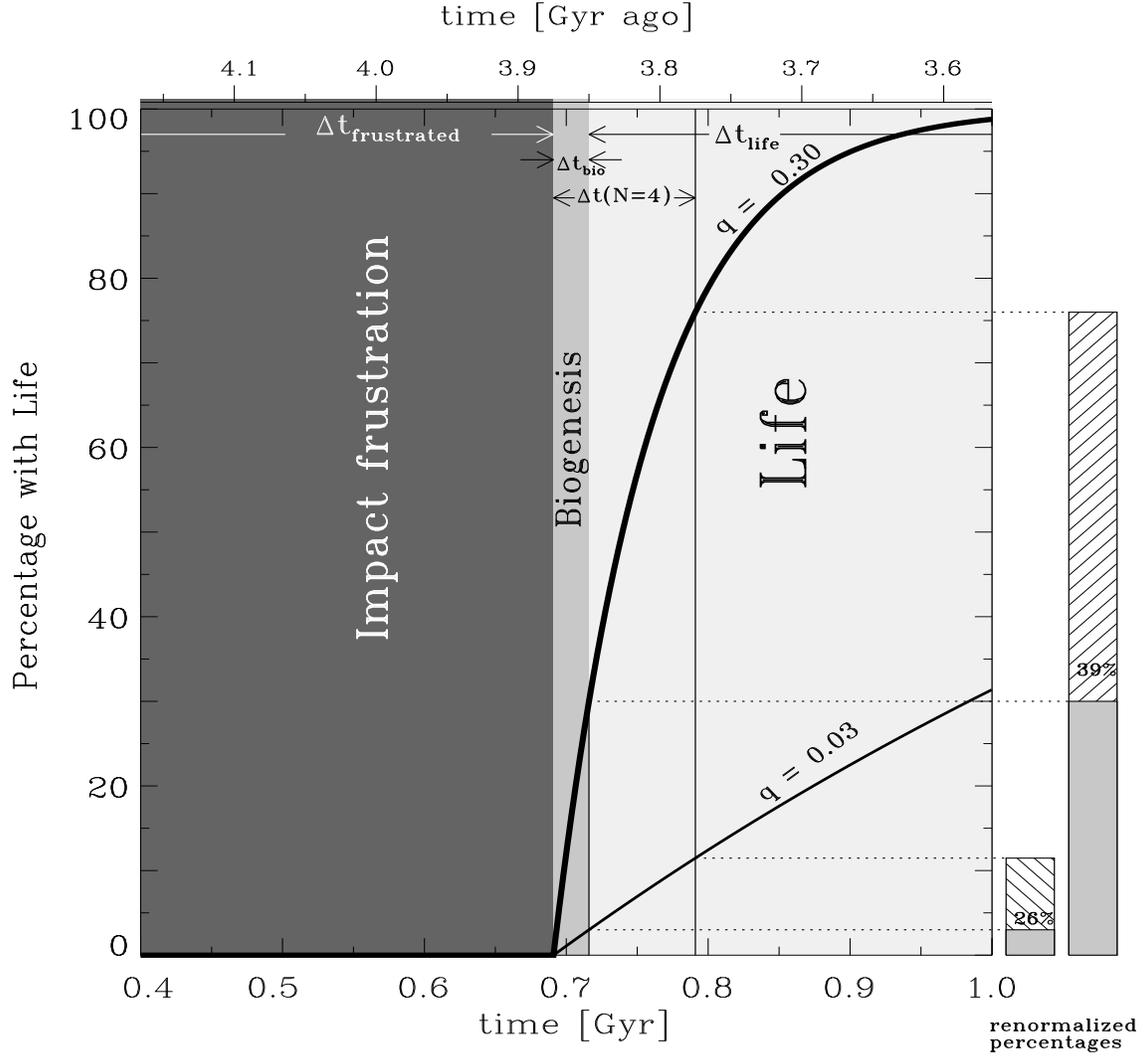,height=14.0cm,width=15.0cm}}
\caption{Zoomed in version of the previous figure with two 
differences: 1) The window for biogenesis  shown here is at the short end 
of the range permitted by observations: $\tb = 25$ Myr.
2) In the previous figure, by normalizing the histograms to 4.566 Gyr,
we have assumed that
Earth is a random member of a group of planets at least $4.566$ Gyr old upon
which biogenesis could have happened {\bf anytime}
up to $4.566$ after formation, including one million years ago. This ignores the 
non-observability-of-recent-biogenesis selection effect 
(Section \ref{sec:nonobservability}). In this figure, to 
minimize the influence of this selection effect, we only allow biogenesis to occur anytime 
earlier than $3.77$ Gyr ago, i.e., within `$\Delta t(N=4)$'. 
%This time was chosen because $\tf + 4  \times 25$ Myr corresponds 
%to $3.77$ Gyr ago.
The influence of this normalization time $\tn$
is illustrated in Fig. 4 and discussed in Section \ref{sec:nonobservability}}
\label{fig:time4}
\end{figure*}
%%%%%%%%%%%%%%%%%%%%%%%%%%%%%%%%%%%%%%%%%%%%%%%%%
%\clearpage

%%%%%%%%%%%%%%%%%%%%%%%%%%%%%%%%%%%%%%%%%%%%%%%%%%%%%%%%%%%%%%%%%%%%%%%%%
%\clearpage
%\vspace{15cm}
\begin{figure*}[!h!t]
\centerline{\psfig{figure=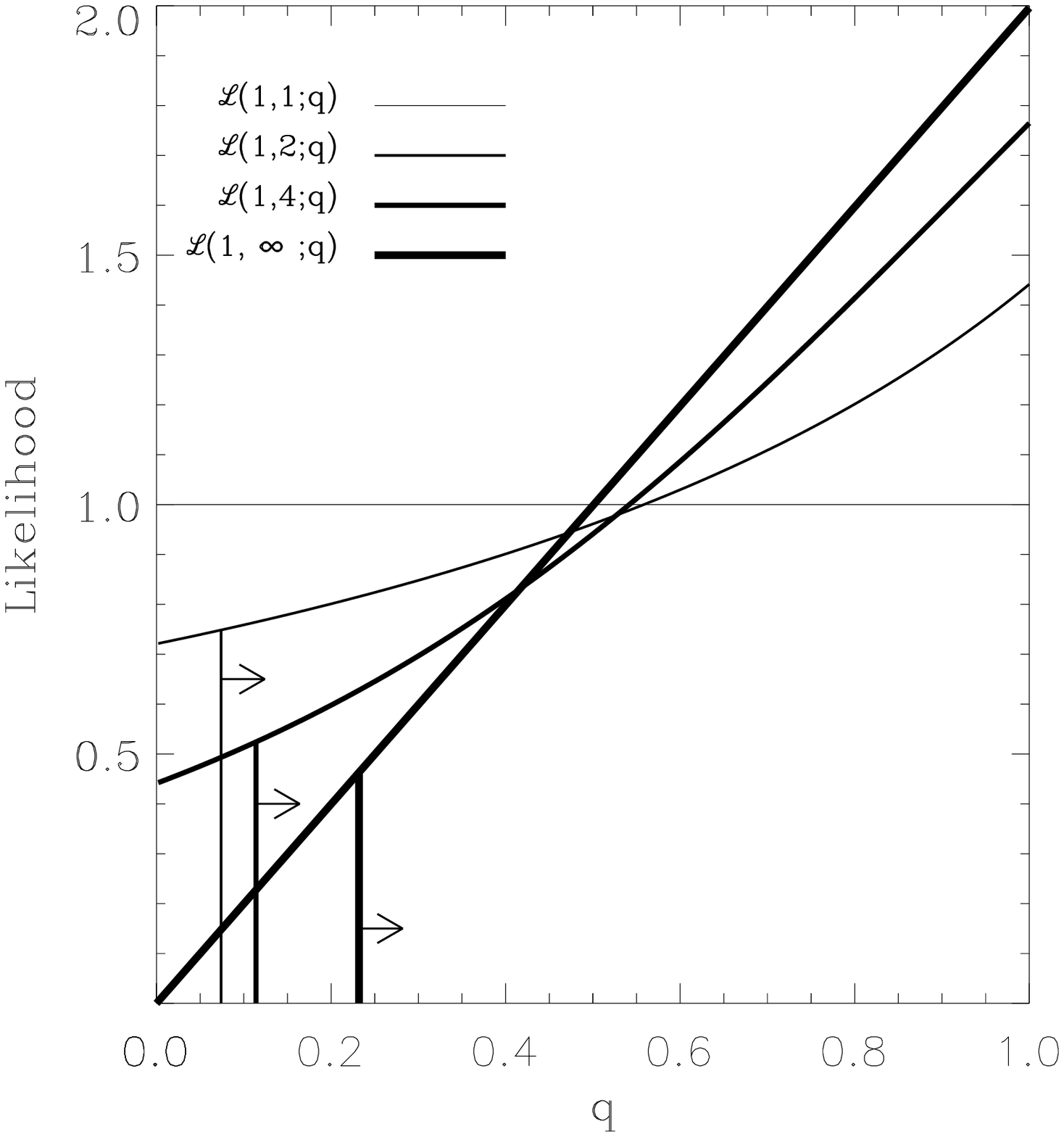,height=14.0cm,width=14.0cm}}
\caption{The effect on the likelihood of varying the 
somewhat arbitrary number of days that the gamblers have to have 
won by to be in the group.
In contrast to  Fig. 1, all the likelihoods of $q$ here are based 
on the information that a gambler (chosen at random from the group 
whose members have won within N days) has won on the first day. 
These likelihoods are plotted and labeled ``${\mathcal{L}}(1,N;q)$'' 
with $N \in \{1,2,4,\infty\}$.
In the biogenesis lottery (just as in the daily lottery) $N$ defines the group and
is a measure of the duration biogenesis could have taken
($\tn = N \times \tb$).
For example, if $N=1$ we can not say anything about
$q$ since the likelihood ${\mathcal{L}}(1,1;q)$ is flat (we have conditioned on `rapid' biogenesis).
While if $N = \infty$ we can put the strongest constraint on $q$. 
If $n=1$ (when it could have been much larger, $1 \leq n \leq N$), we have more
information about $q$ and the likelihood for large $q$ is higher (see Appendix, Eq. \ref{eq:A3}).
The ratio of the probabilities in the histograms in Fig. 3
corresponds to the ratio of the ${\mathcal{L}}(1,4;q)$ likelihoods here:
$\frac{39\%}{26\%} = \frac{{\mathcal{L}}(1,4;q=0.30)}{{\mathcal{L}}(1,4;q=0.03)} = \frac{0.69}{0.46} 
= 1.5$.
That is, with $n=1$ and $N=4$, values of $q \approx 0.30$ are $50\%$ more likely to be 
the case than $q \approx 0.03$ (in Fig. 3, $\tn = 4 \times 25$ Myr).
Notice that for $2 \leq N \leq \infty$, 
the $95\%$ lower limit for $q$ varies only between $0.07$ and $0.23$.
}
\label{fig:nn}
\end{figure*}
%%%%%%%%%%%%%%%%%%%%%%%%%%%%%%%%%%%%%%%%%%%%%%%%%

\section{Selection Effects}
\label{sec:caveats}
\subsection{Daily Lottery $\rightarrow$ Biogenesis Lottery}

In Section 1 we drew parallels between a daily lottery and a 
biogenesis lottery.
Explicitly, these parallels are:
\begin{itemize}
\item
The first day of the lottery corresponds to the end of $\tf$, the 
time when conditions become clement enough for biogenesis.
\item 
All the gamblers had the same (but unknown) chance of winning 
the lottery each time they bought a ticket ($q$ is the probability of 
winning per day). This corresponds to a group of terrestrial planets 
with approximately the same (but unknown) chance of biogenesis as the Earth 
($q$ is the probability of biogenesis within a period of time called here $\tb$).
\item 
We selected a gambler at random from those who had won on or before 
the $N$th day. Thus, we conditioned on winning before a certain time. 
This corresponds to assuming that the Earth is a random member of 
the group of terrestrial planets that has had biogenesis 
on or before the end of $\tn =  N \times \tb$.
Conditioning on biogenesis before this time corresponds to 
correcting for the selection effect that biogenesis had to have 
occurred for us to be here.
\item
A gambler chosen at random, from the group that has won within $N$ 
days, found that he had won on the first day ($n=1$). This corresponds 
to finding that biogenesis has occurred rapidly on Earth, that is,
within $\tb$.
\item
If we set $N=1$, see Fig. 4, ${\mathcal{L}}(1,1;q)$, we are conditioning 
on rapid biogenesis. We are considering a group, all of whose members 
have had rapid biogenesis.  
In this case, a random member having rapid biogenesis can 
tell us nothing about the probability $q$. 
To infer something about $q$ we must have $N > 1$.
\end{itemize}

%%%%%%%%%%%%%%%%%%%%%%%%%%%%%%%%%%%%%%%%%%%%%%%%%%%%%%%%%%%%%%%%%%%%%%%%%
%\clearpage
%\vspace{15cm}
\begin{figure*}[!h!t]
%\centerline{\psfig{figure=lineweaver_fig5.eps,height=13.0cm,width=14.0cm}}
%\centerline{\psfig{figure=lineweaver_newfig5.eps,height=13.0cm,width=14.0cm}}
\centerline{\psfig{figure=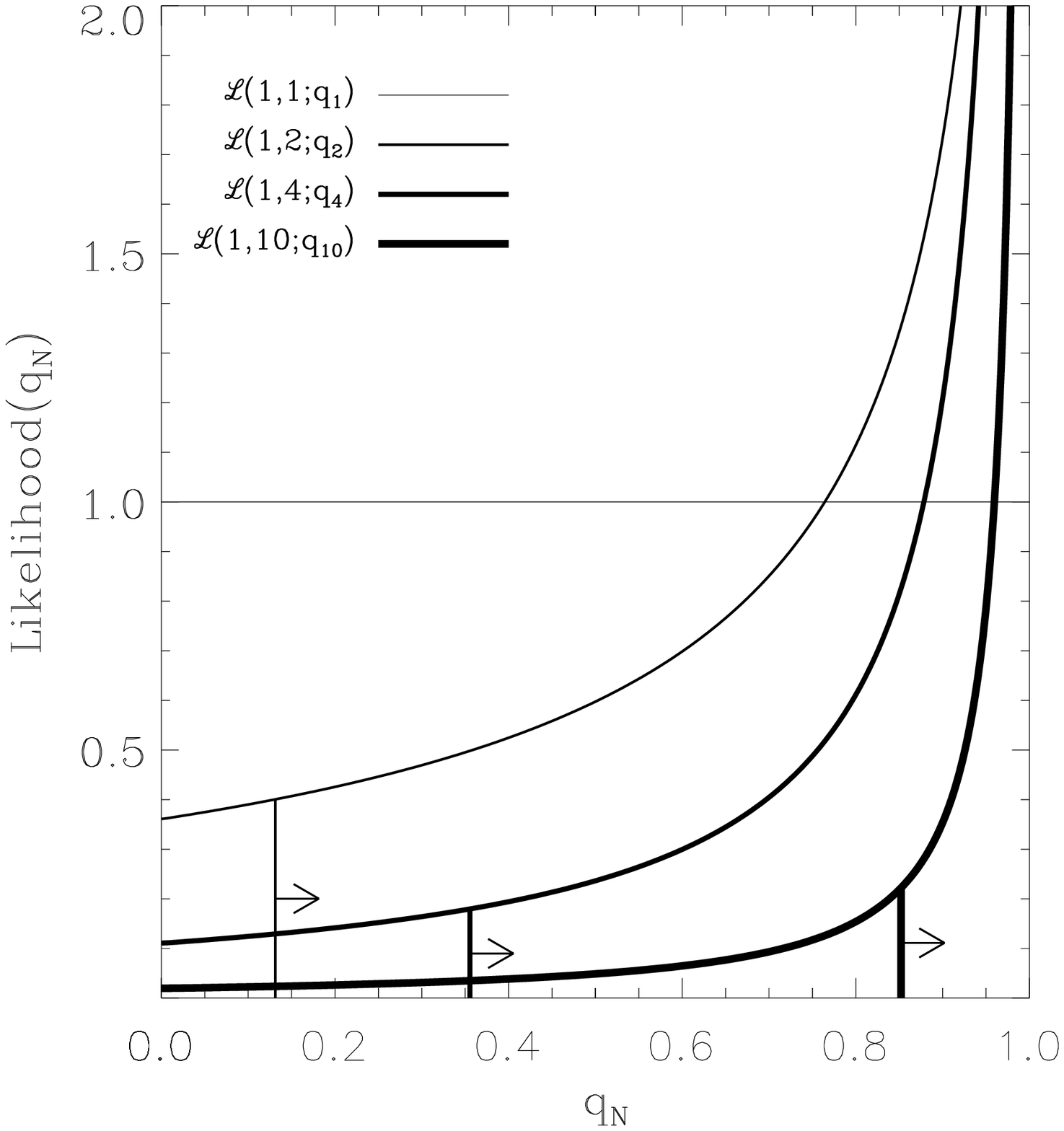,height=13.0cm,width=14.0cm}}
\caption{Likelihood of $q_{N}$.
In Figs. 1 and 4 the likelihood of $q$ is shown, where
$q$ is the probability of winning on any one day.
%$\tb$.
Here we show likelihoods of $q_{N}$,
the unknown probability of winning on or before the $N$th day.
This figure shows the effect of varying $N$.
The information used to compute these likelihoods is that a gambler 
chosen at random from the group whose members have won within $N$ 
days has won on the first day (see Appendix, Eqs. \ref{eq:A5}, 
\ref{eq:A6}, \ref{eq:A7}). Translated this becomes; a planet chosen 
at random, from the group of planets that has had biogenesis 
within $\tn = N \times \tb$, has had biogenesis within $\tb$.
As in the previous figure, if $N=1$ we can say nothing meaningful 
about $q_{1}\: (=q)$.
However, even if $N=2$, we can make a stronger 
statement about $q_{2}$ than we could about $q$: 
$q_{2} > 0.13$ at the $95\%$ confidence level.
The $95\%$ lower limit on $q_{N}$ increases dramatically as $N$ increases, 
constraining $q_{N}$ to be close to $1$.
The ability to extract a useful constraint even if $N$ is low reduces
the influence of the selection effects discussed in Section 
\ref{sec:caveats}.
}
\label{fig:qn}
\end{figure*}
%%%%%%%%%%%%%%%%%%%%%%%%%%%%%%%%%%%%%%%%%%%%%%%%%

\subsection{Non-observability of recent biogenesis}
\label{sec:nonobservability}

If our conclusions from the daily lottery are to apply to biogenesis on 
terrestrial planets we need to correct for the fact that 
the evolution of an observer takes some time.
How long it takes observing equipment, or complex life
or multicellular eukaryotes to evolve, is difficult to say.  
On Earth it took $\sim 4$ Gyr. 
A limited pace of evolution has prevented us from 
looking back at our own history and seeing that biogenesis
happened last year or even more recently than $\sim 2$ Gyr ago
(we consider a plausible range for the requisite time
elapsed since biogenesis to be between 2 and 10 billion years,
or $\Delta t_{\rm evolve} = 4^{+6}_{-2}$ Gyr).

This selection effect for non-recent biogenesis
is selecting for biogenesis to happen a few billion years before the
present regardless of whether it happened rapidly.
It is not a selection effect for rapid biogenesis since the 
longer it took us to evolve to a point when we could measure the 
age of the Earth, the older the Earth became.
Similarly, if biogenesis took $1$ Gyr longer than it actually did, 
we would currently find the age of the Earth to be 
$5.566$ Gyr ($=4.566 + 1$) old; 
`Why is the Earth 4.566 billion years old?',
`Because it took that long to find it out.'
The generalization of this plausible assumption to the ensemble 
of terrestrial planets is necessary if the likelihoods and 
constraints in Figs. 1, 4 and 5 are to be applicable to the group 
of terrestrial planets.

\subsection{Potential Problems}
\label{sec:sensitive}
Any effect that makes rapid biogenesis a 
prerequisite for life would undermine our inferences for $q$.
For example, although it is usually assumed that the heavy bombardment
inhibited biogenesis, energetic impacts may have set up large
chemical and thermal disequilibria that play some crucial role in biogenesis.
We know so little about the details of the chemical evolution that led to
life that heat pulses and rapid cooling after large impacts may be part 
of the preconditions for biogenesis.
If true, the timescale of biogenesis would be linked to the timescale of
the exponential decay of bombardment and biogenesis would (if it occurred
at all) be necessarily rapid; most extant life in the Universe would 
have rapid biogenesis and little could be inferred about the absolute value 
of $q$ from our sample of one.

In a panspermia scenario, the rapid appearance of life on Earth is explained 
not by rapid terrestrial biogenesis as assumed here, but by the ubiquity of 
the `seeds of life' (e.g. Hoyle and Wickramasinghe 1999).  An analysis in the 
context of a panspermia scenario would be subject to the same observational 
constraints as terrestrial biogenesis and would therefore lead to the same 
inferred probability for the appearance of life on other terrestrial planets.

Another potential problem:
suppose the autocatalytic chemical cycles leading to life are 
exponentially sensitive to some still unknown peculiarity of the 
initial conditions on Earth.
In this case, to have the same $q$ as the Earth, our group of 
terrestrial planets may have to be almost indistinguishable rare clones of Earth.
That is, conditioning on planets identical to Earth, (`same $q$')  
would be conditioning on rapid biogenesis ($N=1$) and would prevent us 
from inferring much about $q$ from the observations of rapid
biogenesis on Earth (see ${\mathcal{L}}(1,1;q)$ in Fig. 4).

This is an example of the more general issue of the status of the 
rapidity of biogenesis on Earth. 
Did it have to be rapid? If we assume it could have been otherwise
then we can infer something about $q$. 
If it had to be that way, we can not. The middle
ground might be the most plausible option: biogenesis did not necessarily
have to happen as rapidly as it did, but (to be consistent with our
existence) it may have had to happen within one or two billion years 
of the Earth's formation. 
If this is true, we need to look carefully at 
the influence of varying 
the somewhat arbitrary and counterfactual duration ($\tn = N \times \tb$), 
that biogenesis {\it could} have taken on Earth.
What values of $\tn$ are 
plausible? and how do they effect the results?
This is done in Figure 4 which quantifies the degree of variability 
one can assume for the 
duration of biogenesis and still have interesting constraints on $q$.
The lower $N$ is, the less variability is assumed. 
For example, if $N = 2$, see ${\mathcal{L}}(1,2;q$),
we are assuming that biogenesis could only have taken as long 
as $2\times \tb$ -- enough variability to be able to say something 
about $q$ but small enough to maintain the non-observability of 
recent biogenesis.

Figure 5 allows us to generalize our inferences about $q$ (the probability of 
biogenesis within $\tb$) to inferences about $q_{N}$ (the probability of 
biogenesis within an arbitrary time, $\tn$).  
Specifically, it shows that we only need to be able to assume that 
biogenesis could have been twice as long as $\tb$ 
to have interesting constraints:
$q_{2} > 0.13$ at the $95\%$ confidence level (see ${\mathcal{L}}(1,2;q_{N}$).
This is the result reported in the abstract.

\section{How Common is Life?}
\label{sec:Drake}
\subsection{Relation of our analysis to the Drake Equation}

The Drake Equation was devised to address the question of 
`How many communicative civilizations are in our Galaxy?' (e.g. Sagan 1973).
It has been criticized as ``a way of compressing a large amount
of ignorance into a small space'' (Oliver, cited in Dicke 1998). 
Despite its short-comings, it continues to focus the efforts of the SETI
community. An important goal of the SETI community is to turn
its subjective probabilities into mathematical probabilities. 
We have done that here for one of the most important terms.

We are interested in a simpler question: 
`How common is {\it life} in the Universe?'
Our question is simpler because life is more generic than 
intelligent or technological life.
We modify the Drake Equation to address our question and introduce
a parameter $F$ which is a measure of how common life is in 
the universe. $F$ is the fraction of stars in our galaxy today, 
orbited by planets that have had independent biogenesis,
\be
F= \frac{N_{l}}{N_{*}} =f_{p}\;f_{e}\; f_{l}
\label{eq:F}
\ee
where,\\
$N_{l}$: number of stars in our Galaxy orbited by planets that have had 
independent biogenesis.\\
$N_{*}$: number of stars in our Galaxy.\\
$f_{p}$: fraction of stars in our Galaxy with planetary systems.\\
$f_{e}$: fraction of these planetary systems that 
have a terrestrial planet suitable for life in the same way as the Earth, that is,
they have approximately the same probability $q$ as the Earth.\\
$f_{l}$: fraction of these suitable planets on which biogenesis has occurred.\\

Many recent observations of the frequency and age 
dependence of circumstellar disks around young stars in star-forming regions
support the widely accepted idea that planet formation is a common by-product 
of star formation and that the fraction of stars with
planetary systems is close to unity, $f_{p} \approx 1$
(e.g. Habing et al. 1999, McCaughrean et al. 2000, Meyer and Beckwith 2000).
Equation \ref{eq:F} then becomes,
\be
F \approx f_{e}f_{l}
\label{eq:Fsimple}
\ee
%In some formulations of the Drake Equation the stellar birthrate is 
%replaced by $R^{*} = N_{*}/t_{gal}$ where $N_{*}$ is the number 
%of suitable long-lived stars in our Galaxy and $t_{gal}$ is the age 
%of our galaxy.
%
If $F > 10^{-2}$ then more than 1\% of all stars  has (or has had) life, 
and we conclude that life is `common' in the Universe.
If $F < 10^{-11}$, we may be the only life 
in the Galaxy and life is `rare'. 

There has been little agreement on the value of $f_{l}$.
Michael Hart (1996) writes ``The value of $f_{l}$ is extremely speculative,'' 
but argues  based on the concatenation of low probabilities that it must 
be extremely small and thus, life elsewhere is improbable.
However Shostak (1998) assumes quite the opposite:
``On the basis of the rapidity with which biology 
blossumed on Earth, we can optimistically speculate that this 
fraction ($f_{l}$) is also one (100\%)''.
Our analysis is a close statistical look at this optimistic speculation.

The relation between the fraction of suitable planets on 
which biogenesis has occurred (`$f_{l}$' in the Drake Equation)
and the $q$ analyzed here is,
\bea
f_{l}(q,N) &=& 1 - (1-q)^{N}  \label{eq:fl1}\\
f_{l}(q,\tn) &=& 1 - (1-q)^{(\frac{\tn}{\tb})}  \label{eq:fl2}\\
f_{l}(q,t) &=& 1 - (1-q)^{(\frac{t-\tf}{\tb})}\;, \label{eq:fl3}
\eea
where Eq. \ref{eq:fl1} refers to the daily lottery 
(see Appendix, Eq. \ref{eq:qN}), 
Eq. \ref{eq:fl2} is the translation to the biogenesis lottery and 
we obtain Eq. \ref{eq:fl3} by using $t= \tf + \tn$.
Equation \ref{eq:fl3} is plotted in Figs. 2 and 3.
Thus, in Eq. \ref{eq:fl3} we have expressed the `$f_{l}$' term of 
the Drake Equation as a function of time, of the observables $\tf$ 
and $\tb$, and of the probability
$q$, for which we have derived the relative likelihood.
In addition, since $f_{l}(q,N) = q_{N}$ (see Eq. \ref{eq:qN}), 
Fig. 5 shows the relative likelihood of $f_{l}$ and  
establishes  quantitative but model-dependent 
(specifically $N>1$ dependent) constraints on $f_{l}$.
For example, for terrestrial planets older than 
$\sim 1$ Gyr ($\sim 2 \times \tb$), 
$f_{l} > 13\%$ at the $95\%$ confidence level.

However, to go from $f_{l}$ to an answer to the question
`How common is life?', i.e., to go from $f_{l}$ to $F$ in 
Eq. \ref{eq:Fsimple}, we need to know $f_{e}$:
the fraction of these planetary systems with a terrestrial planet suitable for 
life in the same way as the Earth.
Our understanding of planet formation is consistent with the idea
that `terrestrial' planet formation is a common feature of star formation 
(Wetherill 1996, Lissauer and Lin 2000).
The mass histogram of detected extrasolar planets peaks at low masses: 
$dN/dM \propto M^{-1}$ and is also consistent with this idea 
(Lineweaver and Grether 2002, Zucker and Mazeh 2001,
Tabachnik and Tremaine 2002).

Our analysis assumed the existence of a group of terrestrial planets with 
approximately the same, but unknown ($q \in [0,1]$)
probability of biogenesis.
It is reasonable to postulate the existence of such a $q$-group 
since, although we do not know the details of the chemical
evolution that led to life, we have some ideas about the factors involved:
energy flux, temperatures, the presence of water, planet orbit, 
residence time in the continuously habitable zone,
mass of the planet, atmospheric composition, 
bombardment rate at the end of planetary accretion and its dependence on
the masses of the large planets in the planetary system, 
metallicity of the prestellar molecular cloud, crust composition, 
vulcanism, basic chemistry, hydration, pH, presence of particular 
clay minerals, amino acids and other molecular
building blocks for life (Lahav 1999). 
Since all or many of these physical variables determine the 
probability of biogenesis, and since 
the Earth does not seem to be special with respect to any of them
(i.e., the Earth probably does not occupy a thinly populated 
region of this multi-dimensional parameter space),
the assumption that the probability of biogenesis on these planets 
would be approximately the same as on Earth is plausible. 
%see however Taylor 2000
This is equivalent to assuming that $q$ is a ``slowly'' varying 
function of environment. If true, $f_{e}$ and $F$ 
are not vanishingly small.
This can be contrasted with the ``exponentially sensitive''
case discussed in Section \ref{sec:sensitive}.

\subsection{Discussion}

Carter (1983) has pointed out that the timescale for the evolution of intelligence on the Earth  ($\sim 5$ Gyr) is 
comparable to the main sequence lifetime of the Sun ($\sim 10$ Gyr).  Under the assumption that these two timescales 
are independent, he argues that this would be unlikely to be observed unless the average timescale for the evolution 
of intelligence on a terrestrial planet is much longer than the main sequence lifetime of the host star 
(see Livio (1999) for an objection to the idea that these two timescales are independent).
Carter's argument is strengthened by recent models of the terrestrial biosphere that indicate that the gradual 
increase of solar luminosity will make Earth uninhabitable in a billion years or so -- several billion years 
before the Sun leaves the main sequence (Rampino and Caldiera 1994, Caldeira and Kasting 1992). 

Our analysis is similar in style to Carter's, however we are concerned with 
the appearance of the earliest life forms, not the appearance of  
intelligent life.
Subject to the caveats raised in Section 3.3 and by Livio (1999), the implications 
of our analysis and Carter's are consistent and complementary: the appearance of life on terrestrial planets may be 
common but the appearance of intelligent life may be rare.

\subsection{Summary}

\begin{itemize}
\item 
Our {\it existence} on Earth does not mean that the probability 
of biogenesis on a terrestrial planet, $q$, is large,  because
if $q$ were infinitesimally small and 
there were only one life-harboring planet in the Universe 
we would, of necessity, find ourselves on that planet.
However, such a scenario would imply either that Earth has a 
unique chemistry or that terrestrial biogenesis has
taken a long time to occur. Neither is supported by the
evidence we have.
Since little can be said about the probability $q$, of 
terrestrial biogenesis from our existence,
we assume maximum ignorance: $0 \leq q \leq 1$.
%
%\item 
We then use the observation of rapid terrestrial biogenesis 
to constrain $q$ (Fig.~4). 
%We find that it is $> 10\%$ at the $95\%$ confidence level. (this is from
% N=4.
%
\item
We convert the constraints on $q$ into constraints on the $f_{l}$ term of
the Drake Equation (the fraction of suitable planets which have life).
For terrestrial planets older than $\sim 1$ Gyr %($\sim 2 \times \tb$), 
we find that $f_{l}$ is most probably close to unity 
and $ > 13\%$ at the $95\%$ confidence level.

\item If terrestrial planets are common and they have approximately
the same probability of biogenesis as the Earth,
our inference of high $q$ (or high $q_{N}$) indicates that
a substantial fraction of terrestrial planets have life
and thus life is common in the Universe.

\end{itemize}
However, there are assumptions and selection effects that complicate
this result:
\begin{itemize}
\item
Although we correct the analysis for the fact that
biogenesis is a prerequisite for our existence,
our result depends on the plausible assumption
that {\bf{\it rapid}} biogenesis is 
not such a prerequisite.

\item
Although we have evidence that the fraction of planets which 
are `terrestrial' in a broad astronomical sense (rocky planet 
in the continuously habitable zone) is large, this may be 
different from the fraction of planets which are
`terrestrial' in a more detailed chemical sense.
Although we can make reasonable estimates of what the crusts 
and atmospheres are made of, without detailed knowledge of the 
steps of chemical evolution, we can not be sure that 
astronomically terrestrial planets have the same $q$ as Earth.
That is, the  fraction of planets belonging to the Earth's $q$-group 
is uncertain. Thus, although we have been able to quantify the $f_{l}$
term of the Drake Equation using rapid biogenesis, 
our knowledge of the $f_{e}$ term is still only qualitative
and inhibits our ability to draw stronger conclusions 
about how common life is in the Universe.
\end{itemize}

{\bf Acknowledgemnets}\\
We acknowledge David Nott and Luis Tenorio for vetting the statistics, Paul Davies, John Leslie and an
anonymous referee for useful comments and Kathleen Ragan for editing.
We thank Don Page for gracefully pointing out an error in the plotting of Eq.~15 in 
Fig.~5 of a previous version. 
C.H.L. is supported by an Australian Research Council Fellowship.
T.M.D. acknowledges an Australian Postgraduate Award.
%%%%%%%%%%%%%%%%%%%%%%%%%%%%%%%%%%%%%%%%%%%%%%%%%%%%%%%%%%%%%%%
\newpage 
\begin{center}
References
\end{center}

\noindent All\`{e}gre, C.J., Manh\`{e}s, G., and G\"{o}pel, C. (1995) The age of the Earth.
{\it Geochim. Cosmochim. Acta} 59, 1445-1456.\\
%
%Barrow, J. \& Tipler, F. 1985,  ``The Cosmological Anthropic Principle'', 
%Section 3.2.\\
%
Caldeira, K. and Kasting, J.F. (1992) The life span of the biosphere revisited. {\it Nature} 360, 721-723. \\
Canup, R.M. and Asphaug, E. (2001) 
Origin of the Moon in a giant impact near the end of the Earth's formation.
{\it Nature} 412, 708-712. \\
%\noindent 
%Beckwith, S.V.W., T. Henning, and Y. Nakagawa 2000.
%Dust Properties and Assembly of Large Particles in Protoplanetary Disks. 
%In {\it Protostars and Planets IV}, (V. Mannings, A.P. Boss and S.S. 
%Russell Eds.)  pp 533-558 Univ. of Arizona Press, Tucson.
%
Carter, B. (1983) The Anthropic Principle and its Implications for Biological Evolution. In  {\it Proc.  R. Soc Discussion
Meeting on the Constants of Physics}, edited by W.H. McRea and M.J. Rees, R. Soc., London and in 
{\it Philos. Trans. R. Soc. London}, A310, 347-355\\
%
%\noindent
%Chang, S. 1994, ``THe planetary setting of prebiotic evolution'' in
%Early Life on Earth ed S. Berstrom, Columbia Univ. Press, (p14)
%
Charnley, S.B., Rodgers, S.D., Kuan, Y-J., and Huang, H-C. (2002) 
Biomolecules in the Interstellar Medium and Comets. {\it Adv. 
Space Res.}, Vol. 30, 6, 1419-1431  astro-ph/0104416.\\
%
%\noindent 
%Chyba, C.F. 1993, Geochim. Cosmochim. Acta, 57, 3351-3358,
%``Late Heavy Bombardment 3.8 as limit to age of life''
%
%\noindent 
%%Chyba, C.F, Whitmire, D.P. \& Reynolds, R. 2000
%%``Planetary Habitability and the Origins of Life''
%%in Protostars and Planets IV, eds V. Mannings, A.P.Boss, S.S. Russell,
%%Univ. Arizona Press, Tucson, Arizona, Section IV.\\
%
%\noindent 
Cronin, J.R. (1989)
Origin of organic compounds in carbonaceous chondrites.
{\it Adv. Space Res.} 9(2):59-64.\\
Dicke, S.J. (1998) {\it Life on Other Worlds: The 20th-Century Extraterrestrial Debate},
Cambridge University Press, Cambridge, pp. 217-218.\\
%\noindent 
%Dickerson, R.E. and Geis, I.
%``Chemisty, Matter and the Universe'' 1976, Benjamin Inc, Menlo Park, California
%
%Gogarten-Boeckels, M., Hilario, E. \& Gogarten, J.P.
%``The effects of heavy meteorite bombardment on the early evolution - 
%the emergence of the three domains of life''
%Origins Life Evolution, Biosphere 25, 251-264 (1995)
%
%
Habing, H.J., Dominik, C., Jourdain de Muizon, M., Kessler, M.F., 
Laureijs, R.J., Leech, K., Metcalfe, L., Salama, A., Siebenmorgen, R., 
and Trams, N.  (1999) 
Disappearance of stellar debris disks around main-sequence stars 
after 400 million years. {\it Nature} 401, 456-458.\\
Halliday, A. N. (2000)
Terrestrial accretion rates and the origin of the Moon. 
{\it Earth, Planet Sci. Lett.} 176, 17-30.\\
Halliday, A. N. (2001) Earth Science: In the beginning... {\it Nature} 409, 144-145.\\
Hart, M. H. (1996) Atmospheric Evolution, the Drake Equation and DNA: Sparse Life in an Infinite
Universe. In {\it Extraterrestrials: Where are they?} 2nd edition, 
edited by B. Zuckerman and M.H. Hart Cambridge University Press, Cambridge, pp. 215-225.\\
Hartmann, W.K. and Davis, D.R. (1975)
Satellite-sized planetesimals and lunar origin. {\it Icarus} 24, 504-515.\\
Hartmann, W.K., Ryder, G., Dones, L., and  Grinspoon, D. (2000)
The time-dependent intense bombardment of the primordial Earth/Moon system.  
In {\it Origin of The Earth and Moon} edited by R.M.
       Canup and K. Righter, University of Arizona Press, Tucson, pp. 493-512.\\
%
%Hayes, J.M., Kaplan I.R. \& Wedeking, K.W. in Earth's Earliest Biosphere; its
%Origin and evolution 93-134 (Princeton Univ. Press, Princeton, NJ, 1983)
%(To accompany Mojzsis et al.)
%
Hoyle, F. and Wickramasinghe, N.C. (1999) 
{\it Astronomical Origins of Life: Steps Towards Panspermia},
Kluwer Academic, Boston.\\
%
%Kasting, J.F. 1996 ``Habitable zones around
%stars: An Update'' Circumstellar Habitable Zones edt Laurance R. Doyle, Proc. of the First
%International Conference, Travis House Publications p 17 -28\\
%
Kasting, J.F., Whitmire, D.P., and Reynolds, R.T. (1993) Habitable zones around
main sequence stars. {\it Icarus}, 101, 108-128.\\
%
%Koerner, D. \& LeVay, S. 2000  ``Here Be Dragons: The Scientific Quest 
%for Extraterrestrial Life'' Oxford University Press
%
%Krumbein, W. E. et al. 1988, Terra Cognita, 8, 227
%
Lahav, N. (1999) {\it Biogenesis: Theories of Life's Origin}, Oxford University Press,
Oxford, UK.\\
Lazcano, A. and Miller, S.L. (1994)  
How long did it take for life to begin and evolve to cyanobacteria?
{\it J. Mol. Evol.} 39, 549-554.\\
%
%Lineweaver, C.H. 1999, 
%A Younger Age for the Universe.
%{\it Science} {\bf 284}, 1503
%
Lineweaver, C. H. (2001) 
An Estimate of the Age Distribution of Terrestrial Planets in the Universe:
Quantifying metallicity as a Selection Effect.
{\it Icarus} 151, 307-313.\\     % June 15
Lineweaver, C.H. and Grether, D. (2002)  
The Observational Case for Jupiter Being a Typical Massive Planet. {\it Astrobiology}
Vol. 2, Number 3, 325-334, astro-ph/0201003.\\
%
%Lissauer, J.J. 1995. 
%Urey Prize Lecture: On the Diversity of Plausible Planetary Systems.
%{\it Icarus} {\bf 114}, 217-236
%
Lissauer, J.J. and Lin, D.N.C. (2000) 
Diversity of Planetary Systems: Formation Scenarios and Unsolved Problems.
In {\it From Extrasolar Planets to Cosmology: The VLT Opening Symposium}
Proceedings of the ESO Symposium held at Antofagasta,
Chile, 1-4 March 1999, edited by J. Bergeron and A. Renzini, Springer-Verlag, Berlin  p. 377.\\
Livio, M. (1999) How rare are extraterrestrial civilizations and 
when did they emerge? {\it Astrophys. J.} 511, 429-431. \\
Maher, K.A. and Stevenson, D.J.  (1988) 
Impact Frustration of the Origin of Life.
{\it Nature} {\bf 331}, 612-614.\\
McCaughrean, M.J., Stapelfeldt, K.R., and Close, L.M. (2000)
High-resolution optical and near-infrared imaging of young circumstellar 
disks. In  {\it Protostars and Planets IV}, edited by V. Mannings, A.P. Boss, and 
S.S. Russell, University of Arizona Press, Tucson, Arizona, pp. 485-507.\\
%
%McKay, et al. 1996, Science?
%
Meyer, M.R. and Beckwith, S.V.W. (2000)
Structure and Evolution of Circumstellar Disks Around Young Stars: New 
Views from ISO. In {\it ISO Surveys of a Dusty Universe} edited by 
D. Lemke, M. Stickel and K. Wilke, Springer-Verlag, Heidelberg pp. 347-355.\\
%
%Meyer, S. 1976, ``Data Analysis for Scientists'', Wiley
%
Miller, S.L. (1982) Prebiotic Synthesis of Organic Compounds in {\it Mineral Deposits and the Evolution of the Biosphere}, edited
by W.D. Holland and M. Schidlowski  Springer-Verlag, Berlin pp. 155-176.\\
Mojzsis, S.J., Arrhenius, G., McKeegan, K.D., Harrison, T.M., Nutman, A.P., 
and Friend, C.R.L. (1996) Evidence for life on Earth before 3800 million years ago.
{\it Nature} 384, 55-59.\\
%
%Mojzsis, S.J. et al., 1999
%``Origin of life from apatite dating? Reply'' Nature, 400, 127
%(In reply to Sano's doubts)
%
%Mojzsis, S.J., Harrison, T.M. Pidgeon, R.T. 2001, Nature, 409, 178-181
%``Oxygen-isotope evidence from ancient zircons for liquid water at the Earth's
%surface 4,300 Myr ago''
%
%Morbidelli et al. 2001, ``Source Region and timescales for the Delivery 
%of Water to Earth''
%
%Morris, S.C. 2001(2) Convergences.
%
Nealson, K.H. and Conrad, P.G. (1999) {\it Life: past, present and future}. 
Phil. Trans. R. Soc. Lond. B 1923-1939.\\
%
%Neukum and Ivanov 1994,
%``Crater size distributions and impact probabilities on Earth
%from Lunar, terrestria-planet and astreoid cratering data''
%in Hazards due to comets and asteroids, T. Gehrels ed Univ. of Arizona Press,
%Tucson, Arizona, 359-416
%Late Heavy Bombardment
%
%Newsom, H.E. \& Taylor, R.R. Nature, 338, 29-34 (1988).
%
%Norris, R. 2000. 
%How old is ET?
%In {\it When SETI Succeeds: The impact of high-information Contact} (A. 
%Tough Ed.), Foundation of the Future, Washington DC
%{\it Acta Astronomica}, IAA.9.1.05, in press
%
%Oberbeck, V.R. and Mancinelli, R.L. 1994, 
%``Asteroid impacts, microbes and the cooling of the atmosphere. 
%BioScience, 44, 174-177
%
Oberbeck, V.R. and Fogleman, G. (1989)
Estimates of the maximum time required to originate life.
{\it Origins of Life and Evolution of the Biosphere} 19, 549-560.\\
Pace, N.R. (1991) Origin of life -- Facing up to the Physical Setting.
{\it  Cell}, 65, 531-533.\\
%
%Patterson, C.C. 1956 ``The Age of Meteorites and the Earth'' Geochim \& Cosmochim. Acta 10, 230-237%
%
Rampino, M.R. and Caldeira, K. (1994) The Goldilocks Problem: Climatic Evolution and Long-Term Habitability of
Terrestrial Planets. In {\it Annual Review of Astronomy and Astrophysics} 
Vol 32 pp. 83-114.\\
%
%Rosing, M.T. 1999,
%Independent of Mojzsis isotopic dates cited by Whitehouse above
%``13C-depleted Carbon Microparticles in > 3700-Ma Sea-Floor Sedimentary Rocks
%from West Greenland.'' Science, January 29, 1999, 283, 674-676
%
Sagan, C. (1973) {\it Communication with Extraterrestrial Intelligence}
MIT Press, Cambridge. MA.\\
%
%Sano, Y. et al. 1999, 
%``Origin of life from apatite dating? Comment'' Nature, 400, 127
%For doubts on Mojzsis  paper see
%%
%
%Schellinhuber, H.-J. PIK Potsdam,
%Life Windows of planets, probability of planets
%
%Schopf, J.W. 1983, on stromatilite mineralization in 
%``Earth's Earliest Biosphere: its Origin and Evolution''
%Princeton Univ. Press pp 543 
%
%Activation energies in complex molecular systems (10's of kcal/mol)
%Journal of Chemical Physics, Aug. 1, 1996, 105, 5, 1902-1921
%
%Schopf, J.W. 1993, 
%``Microfossils in the early Archean Apex Chert. New evidence for the antiquity
%of life'' Science, 260, 640-646
%See doubts on Schopf stuff in Edinburgh.
%
Shostak, S. (1998) {\it Sharing the Universe}, Lansdowne, Sydney  p. 180.\\
%
%Simpson, G.G. 1976, in Goldsmith, ET readings 'On the non-prevalence of homonoids'
%
Sleep, N.H., Zahnle, K.J., Kasting, J.F., and Morowitz, H.J. (1989)
Annihilation of ecosystems by large asteroid impacts on the early Earth.
{\it Nature} 342, 139-142.\\
Sleep, N.H., Zahnle, K., and Neuhoff, P.S. (2001)
Initiation of clement surface conditions on the early Earth. 
{\it Proc. Nat. Acad. Sci.} USA, 98, 3666-3672.\\
%
%Sleep, N.H. and Zahnle, K. 1998
%``Refugia from asteroid impacts on early Mars and the early Earth''
%J. Geophys. Res. 103, 28529-28544
%
Stetter, K.O. (1996) Hyperthermophiles in the history of life.
In {\it Evolution of the Hydrothermal Ecosystems on 
Earth (and Mars?)} Ciba Foundation Symposium 202 edited by
G.R. Bock and J.A. Goode, John Wiley \& Sons, Chicester, England
 pp. 1-10.\\
%
%Sugitani, K. Precambrian Res. 57, 21-47 (1992)
%(supports Schopf)
%
%Taylor, S.R. 1999, ``On the difficulties of making Earth-like planets''
%Meteoritics \& Planetary Science 34, 317-329
%
Tabachnik, S. and Tremaine, S. (2002) Maximum-likelihood method for estimating
the mass and period distributions of extra-solar planets. {\it MNRAS}
335, 1, 151-158, astro-ph/0107482.\\
%
%Turcotte, 1990, Earth and Planetary Science Letters, 48, 53-58
%
%
Trimble, V. (1997)
Origin of the Biologically Important Elements.
{\it Origins of Life Evol. Biosphere}  27, 3-21.\\   
Wetherill, G.W. (1996) 
The Formation and Habitability of Extra-Solar Planets.
{\it Icarus} 119, 219-238.\\
%
%Whitehouse, M. 2000,
%``Time constraints on when life began: The oldest record of life on Earth''
%The Geochemical News, Quarterly Newsletter of The Geochemical Society, 103,10-14
%April 2000.
%( for doubts on Mojzsis  paper)
%
%Whittet, D. 1997. 
%Galactic metallicity and the origin of planets (and life).
%{\it Astron. and Geophys.} {\bf 38} (5), 8
%
%Wickramasinghe, N.C. http://www.panspermia.org/balloon2.html
%
%Wilde, S.A. et al. 2001,``Evidence from detrital zircons for the existence of 
%continental crust and oceans on the Earth 4.4 Gyr ago'' Nature, 409, 175-178.
%
%Zahnle, K.J, Kasting, J.F. and Pollack, J.B. 1988, ``Evolution of a steam
%atmosphere during Earth's accretion'' Icarus, 74, 62-97
%
%Zahnle, K.J. and Sleep, N.H. 1997,
%``Impacts and the early evolution of life'' in Comets and the origin and evolution
%of life, P.J.Thomas, C.F.Chyba dn C.P.McKay eds, 175-208, Springer-Verlag, New York
Zucker, S. and Mazeh, T.  (2001) Derivation of the Mass Distribution of Extrasolar Planets with
MAXLIMA- a Maximum Likelihood Algorithm. {\it Astrophys. J.} 562, 1038-1044, astro-ph/0106042.\\

\clearpage
\section{Appendix: Likelihood Computations}
Let the unknown but constant probability of winning a daily lottery be $q$.
Given the information that a gambler who buys one ticket each day
for $n$ days, lost on the first $n-1$ days and
won on the $n$th day, we can compute the likelihood function for $q$
(probability of the data, given the model $q$):
\be
L(n;q) = (1-q)^{n-1}q.   
\label{eq:A1}
\ee
This is equal to the fraction of all gamblers who first experienced ($n-1$) losses 
and then won.
Given only the information that the gambler won at least once on or 
before the $n$th day, the likelihood function for $q$ is:
\be
L(\leq n;q) =  1 - (1-q)^{n}.
\label{eq:A2}
\ee
This is equal to the fraction of all gamblers who have won at least once on or
before the $n$th day.
Given the information that a group of gamblers have won at least once 
on or before the $N$th day, and that a gambler chosen at random from 
this group won at least once on or before the $n$th day ($n \leq N$),
the likelihood of $q$ is:
\be
L(\leq n,\leq N;q)=\frac{L(\leq n;q)}{L(\leq N,q)}=\frac{1-(1-q)^{n}}{1-(1-q)^{N}}
\label{eq:A3}
\ee
Out of all the gamblers who have won on or before the $N$th day, this is the fraction
who have won on or before the $n$th day.
Notice that as $N \rightarrow n$ the likelihood of low $q$
increases and that if $N=n$ the likelihood is the same for all $q$
(see Fig. 4).
As $N \rightarrow  \infty$ we have 
$L(\leq n,\leq N;q) \rightarrow L(\leq n;q)$ which yields the tightest 
constraints on $q$.
A normalized likelihood  ${\mathcal{L}}$ (or probability density) is 
defined such that 
$\int_{0}^{1}{\mathcal{L}}(q) dq = 1$. Thus the renormalization conversion is,
\be
{\mathcal{L}}(x) = \frac{L(x)}{\int_{0}^{1}L(x) dx}.
\label{eq:A4}
\ee
The normalized likelihoods for Eqs. \ref{eq:A1}, \ref{eq:A2} and 
\ref{eq:A3} are plotted in  Fig. 1 for the cases $n=3$ and $N=12$. 
The $95\%$ confidence levels cited are Bayesian credible intervals
based on a uniform prior for $q$.

Although $q$ is the probability of winning the lottery in one day, we would
like to generalize and ask what is the probability $q_{N}$ 
of winning the lottery within $N$ days.
In the analogous biogenesis lottery, $q$ is the probability of biogenesis 
on a terrestrial planet with the same unknown probability of
biogenesis as the Earth. The time window for biogenesis constrained by 
observations on Earth, $\tb$, corresponds to one day for the gambler.
However we would like to compute the likelihood of biogenesis after an 
arbitrary period of time $\tn = N \times \Delta t_{biogenesis}$.
Let $q_{N}$ be the probability of winning on or before the $N$th day:
\be
q_{N} = 1 - (1-q)^{N} = L(\leq N,q),
\label{eq:qN}
\ee
(see Eqs. \ref{eq:A2} and \ref{eq:A3}).
We would like to know the likelihood of $q_{N}$ rather than limit 
ourselves to the likelihood of $q\; ( = q_{1}$).
Suppose the information is the same that was available to compute 
$L(\leq n, \leq N;q)$.
That information  is:  a randomly chosen gambler from 
the group of gamblers who have won after $N$ days, won on or before the $n$th day.
We want $L(\leq n, \leq N;q_{N})$.
The relationship between the likelihood of $q_{N}$ and the likelihood of $q$ is
\be
{\mathcal{L}}(\leq n, \leq N;q_{N})dq_{N} = {\mathcal{L}}(\leq n, \leq N; q)dq
\label{eq:A5}
\ee
which, with $dq_{N}/dq = N(1-q)^{N-1}$ (from Eq. \ref{eq:qN}) becomes
\bea
{\mathcal{L}}(\leq n, \leq N;q_{N}) &=& \frac{{\mathcal{L}}(\leq n, \leq N;q)}{dq_{N}/dq} \label{eq:A6}\\
&=& \frac{{\mathcal{L}}(\leq n, \leq N ;q)}{N(1-q)^{N-1}},
\label{eq:A7}
\eea
which is plotted in Fig. 5 and has, as expected, relatively larger
likelihoods for larger values of $q_{N}$.

%If $q_{N} = 1 - (1-q)^N$ then  $q  = 1 - (1-q_{N})^1/N$ then we have
%\bea
%L( \leq n, \leq N;q) &=& L( \leq n, \leq N;q(q_{N})\\
%                      &=&\frac{1-(1-(1-(1-q_N)^1/N))^n}{q_{N}}\\
%                      &=&\frac{1-(1-q_N)^n/N}{q_{N}}
%\eea

%The lower and upper $x\%$ confidence levels ($q_{lo}$ and $q_{hi}$) are 
%defined by two intergrals:\\
%$(1-x)/2 = \frac{\int_{0}^{q_{lo}} PDF(q)\;dq   }{   \int_{0}^{1} P(q)\;dq   }$\\
%and\\
%$(1-x)/2 = \frac{ \int_{q_{hi}}^{1} P(q)\;dq   }{   \int_{0}^{1} P(q)\;dq   }$\\
%(noticd that the denominator is unnecessary if the PDF is in the numerator 
%rather than just $P$. however this does not include conditioning on 
%winners only. To condition on winners only after $N$ days we need to  
%replace the lower limit of integration $0$ by  some minimal $q$ which, 
%for values lower than q there are no winners.???

\end{document}